\documentclass[twoside]{article}
\usepackage{fleqn,espcrc2}
\usepackage{graphicx}
\usepackage[figuresright]{rotating}

\newcommand{\AmS}{{\protect\the\textfont2
  A\kern-.1667em\lower.5ex\hbox{M}\kern-.125emS}}
\title{
   New approaches  to the determination of the total cross section         }
\author{O.V. Selyugin \address{BLTPh,
        JINR, 141980 Dubna,  Russia }%
	\thanks{ e-mail: selugin@thsun1.jinr.ru}}
\newcommand{\be}{\begin{equation}}
\newcommand{\ee}{\end{equation}}
\newcommand{\ba}{\begin{eqnarray}}
\newcommand{\ea}{\end{eqnarray}}
\newcommand{\baa}{\begin{eqnarray*}}
\newcommand{\eaa}{\end{eqnarray*}}
\newcommand{\bb}{\begin{thebibliography}}
\newcommand{\eb}{\end{thebibliography}}
\newcommand{\ci}[1]{\cite{#1}}
\newcommand{\lab}[1]{\label{#1}}
\newcommand{\stt}{\sigma_{tr}    }

\begin{document}

\begin{abstract}
                New methods have been developed for extracting
      the parameters of diffraction scattering amplitude,
     $\sigma_{tot}$ and $\rho = Re F_N / Im F_N$,
     from  experimental data. The latter determines these parameters
     with less errors and without the knowledge of the normalization
     coefficient.
        The impact of additional information from the measurement
     of the spin-dependence cross section on the determination of the basic
     characteristic of elastic scattering amplitude is examined.
     The new form of the connection between some characteristics
     of the analyzing power and magnitude of the $pp-$ total cross section
     is presented.
 \vspace{1pc}
\end{abstract}

\maketitle

\section{Introduction}

     In reality, in experiment one measure $dN/dt$, as a result of which
      "experimental" data such as $\sigma_{tot}$, slope-$B$, $\rho$
  are extracted
     from  $dN/dt$ with some model asumptions.
  Some approaches are needed to extrapolate
 the measured quantities to $t = 0$. The elastic scattering data
  were constrained by several conditions:
   the imaginary part of the nuclear amplitude has an exponential form
  in the small $t$ region;
  the real and imaginary parts of the nuclear amplitude have the same
     $t$ dependence;
   spin contributions are neglected.
  Note that the value of $\rho$ heavily correlats with the normalization
  of $d\sigma/dt$. Its magnitude weakly impacts  the determination of
  $\sigma_{tot}$ only in the case when the normalization
  is known  exactly.

   There is no any experiment on measurement of the values separately.
    But in some experiments, to reduce experimental errors,
    the magnitude
    of some quantities is taken from another experiment.
   Sometimes, it leads to
    contradiction between the basic parameters for one energy  (for
    example, if we  calculate the imaginary part of scattering
     amplitude, we can obtain a non-exponential behavior). It can lead to
      some errors in the analysis based on the dispersion relations.

The procedure of
extrapolation of the imaginary part of the scattering amplitude is
significant for determining $\sigma_{tot}$.
     The importance of the extrapolated contribution is seen from
 \ci{carb}
where the contribution to $\sigma_{tot}$ of
$\sigma_{obs}$, the directly measured value, and of $\Delta \sigma_{el}$ and
$\Delta \sigma_{inel}$, the extrapolated contributions of the elastic and
inelastic cross sections, are shown at energies $\sqrt s = 30.6, 52.8 $
and $62.7\ GeV$. One can see that the growth of the total cross
sections is due to $\Delta \sigma_{el}$ by $50\%$ for $p p$ and
nearly by $100\%$ for  $p \bar p$ scattering.

  The  analysis of experimental data also
 shows a possible manifestation of spin-flip amplitudes
at high energies \ci{sel}.
The research into  spin effects
will be a crucial stone for different models and will help us
to understand the interaction and structure of particles, especially at large
distances.
All this raises the question about the measure of spin effects
in the elastic hadron scattering at small angles  at
future accelerators.
Especially, we would like to note
the programs at RHIC
where the polarization of both the collider beams will be constructed.

     To obtain the magnitude of $\rho$,
  we fit the differential cross sections either taking into account
 the value of $\sigma_{tot}$ from another experiment, as made
 by the UA4/2 Collaboration, or taking  $\sigma_{tot}$ as a free
 parameter.
 If one does not take the normalization coefficient as a free parameter in
 the fitting procedure, his determination requires the knowledge of
 the behavior of the imaginary and real parts of the scattering amplitude
 in the range of small transfer momenta and the magnitudes of
 $\sigma_{tot}$ and $\rho$.

\section{The method of changing the sign}
    The differential cross sections measured in the experiment
 are described by the squared scattering amplitude.
  From the equation for the differential cross section one can obtain the equation for $\rho$
   for every experimental point - $t_i$ \cite{selbl}
\ba
  \rho(s,t_i) =  \frac{1}{ImF_N(s,t_i)} \ \{ ReF_c(s,t_i) +
                                         \hspace{1cm}  \nonumber \\
  \ [ \frac{1}{\pi} \frac{d \sigma_{i} }{dt} n
     - ( ImF_{c}(s,t_{i}) + ImF_{N}(s,t_{i}) )^2 ]^{1/2} \} .
                                                                 \lab{rsq}
\ea
 As the imaginary part of scattering amplitude is defined by
$  ImF_N(s,t) = H exp(B/2 t)$
where $H=\sigma_{tot}/(4 \pi *0.389)$
 it is evident from  (\ref{rsq}) that the real part depends on
 $n, \sigma_{tot}, B$.

  For the proton-proton scattering this equation has a remarkable property.
 If we expand the expression under the radical sign, we obtain
\be
 (n-1)(ImF_c+ImF_N)^2 + n (ReF_c+ReF_N)^2 .
\ee
 As the real part of the Coulomb scattering amplitude is negative and
 the real part of the nucleon scattering amplitude is positive, it is clear
 that this expression will have a minimum situated on the scale
 of $t$ independent of $n$ and $\sigma_{tot}$.
 As we know the
 Coulomb amplitude, we estimate the real part of the proton-proton
 scattering amplitude at this point. Note that all other methods give us
 the real part only in a sufficiently wide interval of the transfer
 momenta.

  This method works only in the case of the positive real part
 of the nucleon amplitude and it is especially
  good in the case of large $\rho$. So, it is
 interesting for the future experiment at RHIC.

\section{The differential method ("tail of ghost")}
  Let us examine the simplest gedanken case of $p p$-scattering
  and try to determine the sign of  $d \rho(t)/dt$  depending on
  the difference of normalization $n$ and $n + \delta n$ and
   $\delta \sigma =\sigma_{true} - \sigma_{tot-prob}$.
  There,  $\sigma_{true} $ is the magnitude of the true total
  cross section and $\sigma_{tot-prob}$ is the magnitude of the total
  cross section which we take as our first approximation.

  In first, let us suppose that in experiment we find true normalization
  of data $dN/dt$ (so $n=1$) and try to find the derivate of the
  calculated $\rho$ on $t$ ($d \rho/dt$)
   with a small deviation of $n$ ($\delta n$) \cite{selbl}.
 Write for that  case our equation for $\rho$ for two points:
  $ t_1$, where $-F_c = 2 Im F_h$,  and for $t_2$, where $-F_c= Im F_h$;
  and calculate $\Delta \rho = \rho(t_2) - \rho(t_1)$
   for normalization $n + \delta n$ with $\delta n << 1$.
  So, for example, for $t_2$ we can input in (1)
\ba
  \frac{1}{\pi}\frac{d\sigma}{dt}(n+\delta n) =
  2 ImF_{h}^2 (1-\rho+\rho^2) (1+\delta n)
\ea
and obtain, if we take $\sigma_{t-prob}= \sigma_{tr}$,
 that the difference between the calculated $\rho_{i}$ will be
\ba
  \Delta \rho(t_2,t_1) \simeq  \delta n /4
\ea
  and is a function of $t$.
   It is clear that if we find the true magnitude of the total cross
   section and $\delta n =0$, then $\Delta \rho(t_2 , t_1) $ will be equal
   to zero.
                   The same calculations are carried out for $\delta \stt$
  with $\delta n =0$ and show that
    the sign of $\Delta \rho(t_2,t_1)$ depends on
  the sign of $\Delta \sigma_{tot}$.
     Then, if we will calculate  $\Delta \rho(t_2,t_1)$ with differences
  $\delta n$ and $\sigma_{tot}$ we can determine the true magnitude
  of $\sigma_{tot}$.

\section{ Connection between $A_N$ and $\sigma_{tot}$ }
    Let us examine the future  $pp2pp$ Experiment at $\sqrt{s} =500 \ GeV$,
    as an example.
 Elastic differential cross section will be regarded as having $2 \%$
  statistical errors, according to the Proposal of the $PP2PP$ Collaboration.
  Now, in fitting procedure we take into account
  the standard assumption for high energy elastic hadron scattering
  at small angles: the simple exponential behavior
  with slope $B$ of the imaginary and real parts of the scattering amplitude;
  hadron spin-flip amplitude does not exceed $10 \% $ of the hadron spin-non-flip
  amplitude.
   The differential cross section was calculated using
    $ \sigma_{tot} =63.5 \ mb ;   \ \rho=0.15;  \
    B=15.5 \ GeV^{-2} $ ( in variant $I$ with
 $  150 $ points and in variant $II$ with $75$ points)
  from $t_0 = 0.00075$ with $\Delta t = 0.00025 \ GeV^{2}$)
 and
   then put through a special random process using $2 \% $ errors.
   After that, the obtained "experimental" data were fitted.
  The systematic errors were taken into account as the free parameter $n$.
   The result is presented in Table 1 for three ($n=1.$- fixed)
   or four  parameters ($n$ - free).

 \begin{center}

 \begin{tabular}{llllll}

    \multicolumn{5}{l}
   {Table 1. \hspace{5mm} Fit of $d \sigma/dt$  measurements        }  &\\
       \hline
 $N$ & $\sigma_{tot} \ mb $ & $\delta B $ & $\delta \rho$ &$ \delta n$   \\
 $ I$ & $63.54 \pm 0.12$ &  $\pm 0.2$ & $ \pm 0.008$ & fix  \\
 $ I$ & $63.6 \pm 1.25$ &  $\pm 0.3$ & $ \pm 0.02$
                                                 & $ \pm 0.04 $  \\
 $ II$ & $63.5 \pm 0.25$ &  $ \pm 0.7$ & $  \pm 0.01$ & fix  \\
 $ II$ & $64.05 \pm 1.4$ &  $ \pm 1.0$ & $  \pm 0.03$
                                                 & $\pm 0.05 $  \\
  \hline
 \end{tabular}
 \end{center}

  It is clear that the most important value is the coefficient normalization
  of the differential cros section. Its small errors lead to  significant
  errors in the $\sigma_{tot}$.
    So, we see that the normalization of experimental data is the most
  important problem for the determination of $\sigma_{tot}$.

 Lacking better knowledge, we assume that the hadron spin-flip
  amplitude is a slowly varying function of $t$ apart the kinetic factor,
  and
 we parametrize it as
\ba
   \phi^{h}_{5}= \frac{\sqrt{|t|}}{m} (\rho k_2 + i k_1) Im \phi^{h}_{1} ,\lab{sf2}
\ea
  where $\rho$, $k_1$, $k_2$ are slowly changing functions of s.
 The coefficients $k_1$ and $k_2$ are the ratios of the real and
  imaginary parts of the spin-flip to spin-non-flip amplitudes
  without the kinematic factor $\sqrt{|t|}$.
 As a result, the $A_N$ can be written as
\ba
 -  \frac{A_N}{8 \pi P_{B} } \frac{d\sigma}{dt}
   =-Im \phi^{h}_{1} \frac{\alpha}{m \sqrt{|t|}} (\frac{\mu-1}{2}-k_1) \nonumber \\
          + \frac{\sqrt{|t|}}{m} \rho [Im \phi^{h}_{1}]^{2}
	 \Delta k,                                         \lab{kil}
\ea
with $\Delta k = k_2-k_1$.
   We examined also a few variants with different assumptions about the
  magnitudes of the imaginary and real parts of the hadron spin-flip amplitude
  at $\sqrt{s} = 500 \ GeV$.
  Variant $A_i$:  $\phi^{h}_5 =0.$, so $k_1 =k_2 =0$ .
  Variant $B_i$:  $k_1=0.1$, $k_2=0.15$.
  Variant $C_i$: $k_{1} = 0.1$, $k_{2} = -0.15$.
  Variant $C_3$ was made by a random procedure with errors  twice
  smaller, so the errors equal $5 \% \div 10 \% $ (see, Table 2).

 The "experiment" with $10 \% \div 20 \% $ errors
 determine the magnitudes of real and imaginary parts of the hadron spin flip almost
  with $100 \% $ errors. The variant $C_3$, which reflects the experiment
 with $5 \% \div 10 \% $ errors, twice  decreases the errors
   of the magnitude of $\sigma_{tot}$ and
   hadron spin-flip.

  Let us make the fit of both data on the differential cross section and
  the analyzing power. The results are shown in Table 3.
 The added polarization data decrease the error in $\sigma_{tot}$ only
  by $10 \%$ (from $1.25 \ mb$ to $1.1 \ mb$). But the determination of
  the magnitude of the real and imaginary parts of the hadron spin-flip
  amplitude become three time better.
  It is to be noted that the variant $C_3$ leads to decrease of in the error
  of $\sigma_{tot}$  from $\pm 1.25 \ mb $ to $ \pm 0.9 \ mb$.

 \vspace{3mm}

 Table 2.  Fit of $A_N$ \hspace{5mm}

 \vspace{2mm}

 \begin{tabular}{ccccc} \hline
 $N$  & $\sigma_{tot} \ mb $ & $\delta \rho$  & $\delta k_1$ & $ \delta k_2$  \\
 $ A_1$ & $63.5 \pm 3.4$ &  $ \pm 0.08$ & $ \pm 0.07$
                        & $ \pm 0.05$     \\
 $ A_2$ & $63.46 \pm 3.8$ &  $\pm 0.15$ & $ \pm 0.06$
                             & $ \pm 0.1  $   \\
 $ B_1$ & $63.5 \pm 3.8$ &  $ \pm 0.09$ & $ \pm 0.07$
                        & $ \pm 0.11  $ \\
 $ B_2$ & $62.7 \pm 4.$ &  $ \pm 0.3$ & $  \pm 6.3$
                             &$ \pm 5.6  $        \\
 $ C_1 $& $63.4 \pm 3.6$ &  $\pm 0.09$ & $ \pm 0.07$
                        &$ \pm 0.11 $    \\
 $ C_2$ & $63.5 $ -fix &  fix & $ \pm 0.015$
                             &$  \pm 0.011 $   \\
 $ C_3$ & $63.9 \pm 1.83 $ &  $ \pm 0.05 $ & $ \pm 0.037$
                             &$  \pm 0.035 $   \\
  \hline
 \end{tabular}
 \vspace{5mm}

\vspace{2mm}

Table 3.  Fit of $d \sigma /dt$ and $A_N$ (300 points)

 \vspace{2mm}

 \begin{tabular}{cccccc} \hline
 $N$  & $\sigma_{tot} \ mb $   & $\delta \rho$ & $\delta k_1$ & $\delta  k_2$  \\
 $ A$ & $63.4 \pm 1.1$  & $ \pm 0.02$ & $ \pm 0.02$
                        & $ \pm 0.02$      \\
 $ C$ & $63.4 \pm 1.1$  & $ \pm 0.02$ & $ \pm 0.02$
                        & $ \pm 0.03$      \\
 $ C_3$ & $63.2 \pm 0.9$  & $ \pm 0.015$ & $ \pm 0.014$
                        & $ \pm 0.02$      \\
  \hline
 \end{tabular}

 \vspace{3mm}

\section{ Connection between  $t_{max}$ of $A_N$ and $\sigma_{tot}$ }
   As  noted above, most uncertainty in the determination of $\sigma_{tot}$
   using the measurement of $A_N$ came from the error in the beam
   polarization which plays the role of a normalization factor of the
   differential cross section. The point of maximum of $A_N$ is independent
   of the magnitude of the beam polarization. So, it allows us to use
   this value for the extraction of the magnitude of $\sigma_{tot}$.
    Here is the formula \cite{batt} (the case $B1$)
 \ba
 \sigma_{tot}= \frac{9.776 \alpha}{t_{max}}
         [\sqrt{3} - \frac{8}{\mu -1}(\rho I - R )]
                                           - (\rho - \alpha \varphi )],
                         \nonumber
\ea
 where $I=k_1$ and $R= \rho k_2 $, and these coefficients are unknown.
 Its determination will depend on the magnitude of beam
  polarization.
  To reduce the impact of the hadron spin-flip amplitude, it was
  proposed to used the new value -  $t_{max2}$,
   the place of the maximum of $d\sigma /dt A_{N}^2$ \cite{sel-nic} . The derivation of this
  value gives (the case $A1$ )
 \be
       \sigma_{tot} = 8 \pi \  0.39 (mb/GeV^2)  \ \alpha /t_{max2}.
		                     \nonumber
 \ee
  In the case of the exponential behavior of the scattering
  amplitude, one obtains (the case $A2$ )
 \be
    \sigma_{tot} = 9.776 \alpha [1/t_{max2} - B/2]  ,
			                  \nonumber
 \ee
   where $B$ is the slope of the differential cross section.
  In our previous work we showed that it could be obtained from
  the measurement of
  $A_N$ of some ratio of the real and imaginary part of the hadron
  spin-flip amplitude which is independent of the magnitude of
  the beam polarization
$ \rho (k_2-k_1)/(1-k_1) $.
     Using this ratio, we can obtain the relation between $\sigma_{tot}$
   and the point of maximum $A_N$ in the form (the case $C2$)
\ba
  \sigma_{tot}=  9.776 \alpha  ( \frac{1}{ t_{max} } + B)
         [ \frac{ \sqrt{3+4 \rho^2} }{1+\rho^2} \nonumber \\
   \frac{8 \rho \Delta k }{ (\mu-1)(1+2\rho)(1-k_1)}
                                           - (\rho - \alpha \varphi )].
\ea
 The calculation of $\sigma_{tot}$ by using these formulae
 is shown in Table 4 for $\sqrt{s}=52 \ GeV$
 and in  Table 5   for $\sqrt{s}=540 \ GeV$
  for different variants of the magnitude of the hadron spin-flip
 amplitude. For comparison,         the variant $C1$    is show
 without taking into account of the contribution of the hadron spin-flip
 amplitude (the case $C1$)
\ba
  \sigma_{tot}=   \hspace{5cm} \nonumber \\
  9.776 \alpha  ( \frac{1}{ t_{max} } + B)
         [ \frac{ \sqrt{3+4 \rho^2} }{1+\rho^2}
                                           - (\rho - \alpha \varphi )].
\ea

\section{ Conclusions}
   The magnitudes of $\sigma_{tot}$, $\rho$ and slope - $B$ have to be
      determined in one experiment and their magnitudes  depend on
      each other.
   The normalizations of $dN/dt$ and $A_N$ are most important
     for the determination of these values.
     The new methods of extracting  magnitudes of these quantities
       are required.
   Some of such ideas were shown in this talk.
 Additional information on $A_N$ slightly reduces
  the errors in the size of $\sigma_{tot}$.
  The presented new formulae for the determination of the magnitude of
   $\sigma_{tot}$
   using the place of $t_{max}$ of the CNI
   give sufficiently good results in a wide energy region and can be used
   as an  additional method for the determination of $\sigma_{tot}$.

 Table 4. $\sigma_{tot}$ as function of $t_{max}$

 \vspace{2mm}

 \begin{tabular}{llllll} \hline
   \multicolumn{6}{c}{ Input $\rho=0.1$,$ B=13. $,
                    $\sigma_{tot}$ =43.0  mb } \\
$k_1, k_2$ & B1 & A1$t_{m2}$    & C1$t_{m}$ & C2$t_{m}$ \\
 0., 0.   & 42.0  & 43.2  & 43.3  & 43.3   \\
 0.1, 0.2   & 41.9  & 42.1  & 42.2  & 43.3   \\
 0.2, 0.2   & 42.0  & 43.2  & 43.3  & 43.3   \\
 0.0, 0.2   & 42.0  & 41.0  & 42.1  & 43.4   \\
 $\rho=0.075$         &   &  &     &    \\
0.0, 0.2  & 41.9  & 41.4  & 41.7  & 43.5   \\
  \hline
 \end{tabular}

\vspace{5mm}

Table 5. $\sigma_{tot}$ as function of $t_{max}$

\vspace{2mm}

 \begin{tabular}{llllll} \hline
   \multicolumn{6}{c}{ Input $\rho=0.15$,$ B=15.5 $,
                    $\sigma_{tot}$ =63.5  mb } \\
$k_1, k_2$ & B1 & A1$t_{m2}$    & C1$t_{m}$ & C2$t_{m}$ \\
 0., 0.   & 62.1  & 63.5  & 63.3  & 63.5   \\
 0.2, 0.2 & 62.3  & 63.5  & 61.9  & 63.5   \\
 0.0, 0.1 & 62.8  & 61.8  & 61.5  & 63.7   \\
 0.1, 0.2 & 62.0  & 61.3  & 60.7  & 63.3   \\
 0.0, 0.2 & 62.2  & 59.3  & 58.5  & 63.7   \\
 -0.1, -0.2 & 61.0  & 64.9  & 64.9  & 63.1   \\
 $\rho=0.1$         &   &    &   &    \\
0.2, 0.2  & 62.0  & 63.5  & 63.6  & 63.6   \\
 $\rho=0.0 $         &     &  &   &    \\
 0.2, 0.2   & 61.9  & 63.2 & 63.8  & 63.7   \\
  \hline
 \end{tabular}

\vspace{4mm}

  {\it Acknowledgments.}
I would like to express my sincerely thanks to the Organizing
 Committee and especially to R. Fiore and A. Papa for the kind invitation and the financial support
at  such remarkable Conference, and W. Gurin, B. Nicolescu and E. Predazzi
for fruitful discussions.

\end{document}